\documentclass{sig-alternate}
\usepackage[pdftex]{color}

\usepackage{cite}
\usepackage{graphicx}
\usepackage{url}
\usepackage{balance}
\usepackage{xspace}

\newcommand{\eg}{e.\,g.\ }

\newcommand{\impact}[4]{\textsf{\small [#1$\,\vert\,$\uppercase{{\scriptsize#2}}] 
$\stackrel{#4}{\longrightarrow}$ [#3]}\xspace}

\newcommand{\impactp}[3]{\impact{#1}{#2}{#3}{+}}
\newcommand{\impactn}[3]{\impact{#1}{#2}{#3}{-}}

\newcommand{\attribute}[1]{\textsf{\small\uppercase{{\scriptsize#1}}}}

\begin{document}
\conferenceinfo{WoSQ'08,} {May 10, 2008, Leipzig, Germany.}  
\CopyrightYear{2008} 
\crdata{978-1-60558-023-4/08/05}

\title{Managing Quality Requirements Using \\Activity-Based Quality Models}

\numberofauthors{1}

\author{
  \alignauthor Stefan Wagner, Florian Deissenboeck, and Sebastian Winter\\
    \affaddr{Institut f\"ur Informatik}\\
    \affaddr{Technische Universit\"at M\"unchen}\\
    \affaddr{Garching b.\ M\"unchen, Germany}
}

\maketitle

\begin{abstract}
Managing requirements on quality aspects is an important issue in the 
development of software systems. Difficulties arise from expressing them
appropriately what in turn results from the difficulty of the concept of
quality itself. Building and using quality models is an approach to handle
the complexity of software quality. A novel kind of quality models uses the
activities performed on and with the software as an explicit dimension. These
quality models are a well-suited basis for managing quality requirements from
elicitation over refinement to assurance. The paper proposes such an approach
and shows its applicability in an automotive case study.
\end{abstract}





\section{Quality Requirements}

Quality requirements are usually seen as part of the \emph{non-functional}
requirements of a system. Those non-functional requirements describe
properties of the system that are not its primary functionality. ``Think of
these properties as the characteristics or qualities that make the product
attractive, or usable, or fast, or reliable.'' \cite{robertson99} Although
this notion of non-functional requirements is sometimes disputed, there
always exist requirements that relate to specific qualities of the system
\cite{glinz05}. We call those demands \emph{quality requirements}.

\subsection{Problem}
Quality requirements are an often neglected issue in the requirements
engineering of software systems. A main reason is that
those requirements are generally difficult to express in a measurable
way what also makes them difficult to analyse \cite{nuseibeh00}. One reason
probably lies in the fact that quality itself ``[\ldots] is a complex
and multifaceted concept.''~\cite{1984_garvind_product_quality} It
is difficult to assess and thereby also the definition of quality
requirements is a complex task. Especially incorporating the various aspects of
all the stakeholders is often troublesome. Hence, the problem is how 
to elicit and assess
quality requirements in a structured and comprehensive way.

\subsection{Contribution}
We propose a 5-step approach for managing quality requirements using a
two-dimensional quality model \cite{deissenboeck07}. This quality model uses
activities as one dimension and describes the influences of system entities
(and their attributes) on those activities. These two dimensions can
conveniently be used as a structure for quality requirements as well. The
stakeholders define the activities they perform on and with the system. They
provide the most abstract level for quality requirements. Refinements can be
made by analysing which system entities are affected by which activities.
Finally, a direct traceability from the quality assurance to the quality
requirements is given by the quality model. This is the case because the
model can be used as basis for quality assurance techniques such as reviews.

\subsection{Outline}

We start with describing related work in quality modelling and eliciting
and structuring quality requirements in Sec.~\ref{sec:related}.
Sec.~\ref{sec:quality_modelling} introduces activity-based quality models
and their advantages over traditional approaches.
In Sec.~\ref{sec:elicitation}, we propose 
an approach to elicit and
refine quality requirements based on such activity-based quality models.
The relation to assuring the requirements is described in 
Sec.~\ref{sec:assurance}.
Then, in Sec.~\ref{sec:case_study}, the approach is validated in a case study.
Final conclusions are given in Sec.~\ref{sec:conclusions}.

\section{Related Work}
\label{sec:related}

Various approaches for non-functional requirements have been proposed. 
The standard
IEEE Std 830-1998 \cite{ieee98} considers requirements specifications in
general. It concentrates strongly on functional
issues and quality requirements play only a minor role. 
Ebert discusses in \cite{ebert97} an approach for managing non-functional
requirements. He classifies them in \emph{user-oriented} and
\emph{development-oriented} that gives them a first structure. However,
this is still too coarse-grained to be applied fruitfully.
Also more general approaches such as
\cite{cysneiros04,mylopoulos92} do not impose a sufficient structure on the
quality requirements that foster elicitation or assurance.
More structure is provided by the UMD approach \cite{basili04}. However, it
focuses mainly on \emph{issues} that should be avoided and hence do not
provide enough connection to the stakeholders.
Finally, Doerr et al.~\cite{doerr05} included the use of quality models in
their approach to non-functional requirements. However, the used quality
models themselves provide no direct connection to the stakeholders.

\section{Quality Modelling}
\label{sec:quality_modelling}

To be able to efficiently manage quality requirements one needs a means to 
express them in a concise and consistent manner. Since the 1970ies a number of 
\emph{quality models} have been proposed to achieve 
this~\cite{1977_mccallj_quality_factors, 1978_boehmb_software_quality, 
1992_omanp_maintainability, 1995_dromeyr_product_quality, 
2003_iso_standard_9126_1}. However, as is argued in \cite{deissenboeck07} 
these approaches have a number of shortcomings. Most importantly, they fail 
to make explicit the interrelation between system properties and the 
activities carried out on or with the system by the various stakeholders. 
We regard the omission of activities as a serious flaw as the activities 
performed on and with the system
largely determine the overall life-cycle cost of a software system. Moreover, 
the activities provide a natural criterion for the decomposition of the complex 
concept \emph{quality} that many existing approaches lack.

To address these problems, we propose a consequent separation of activities 
and system entities. This separation facilitates the 
identification of sound quality criteria and allows to reason about their 
interdependencies. To illustrate the activity-based quality model we
use the quality attribute \emph{maintainability} that is known
to have major influence on the total life-cycle cost of software systems.

The 1\textsuperscript{st} dimension of the quality model consists of the 
activities carried out on or with the system by the various stakeholders. 
In the case of maintainability the set of relevant activities depends on 
the particular development and maintenance process of the organisation that 
uses the quality model, \eg the IEEE 1219 standard maintenance 
process\,\cite{1998_ieee_standard_1219}. As activities can be conveniently 
structured in activities and related sub-activities, the 
1\textsuperscript{st} dimension of the model actually forms a tree:
the \emph{activities tree}.

The 2\textsuperscript{nd} dimension of the model, the \emph{entities tree} 
describes a decomposition of the \emph{situation}. 
We use the term \emph{situation} here to express that this tree is not 
limited to a description of the software system itself but also describes 
relevant aspects of the organisation that develops the system. This is 
necessary as organisational aspects like development processes and the 
provided infrastructure are known to have a major impact on the expected 
maintenance effort.

\begin{figure}[ht] \includegraphics[width=0.99\linewidth]{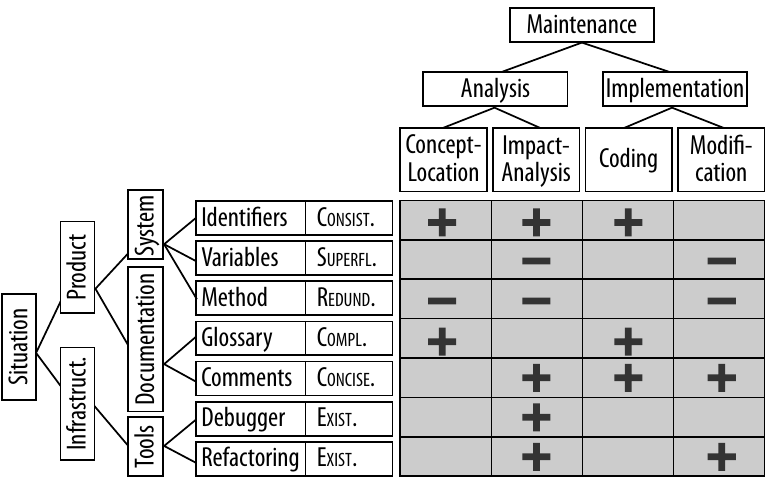} 
\caption{Maintainability model \label{fig:matrix}} \end{figure}

To achieve or measure maintainability in a given project setting we need to 
establish the interrelation between entities and activities. 
This relationship is best
expressed by a matrix as depicted in the simplified Fig.~\ref{fig:matrix}.

As the figure shows, to be able to express these relations, one needs to 
equip the entities with fundamental \emph{attributes} like 
\emph{consistency, completeness, conciseness} or \emph{redundancy}. Now 
entities and activities can be put into relation by the identification of 
\emph{impacts}. An impact is defined as a relation between an entity/attribute
tuple 
and an activity where $+$ expresses a positive and 
$-$ a negative impact.

\begin{center}
\impact{Entity $\mathsf{e}$}{Attribute $\mathsf{a}$}
{Activity $\mathsf{a}$}{+/-}
\end{center}

An example in the figure is \impactp{Identifiers}{conciseness}{Concept 
Location} that expresses that the conciseness of identifier names has a
positive influence on the activity \emph{concept location}. The negative 
impact of 
superfluous variables on the modification of existing code is expressed as 
\impactn{Variables}{super\-fluousness}{Modification}. The example shows 
that even the apparently simple attribute \attribute{existence} can be very powerful 
when we want to state that a proper infrastructure, e.g., a debugger has an 
influence on specific activities.

Such a model is well-suited to classify quality requirements as the 
activities provide a straight-forward relation between the stakeholders, 
that ultimately define quality requirements, and the software system. For 
the example of maintainability, the related stakeholder is the developer. 
His main activity \emph{maintenance} can be broken down 
in more tangible subactivities. This subactivities can then be  related to 
situation entities via basic attributes. In \cite{winter:interactive07} it is 
presented how such a model can be used for the quality attribute 
\emph{usability}. The central stakeholder is the user and his core 
activity \emph{usage} can be decomposed in more specific subactivities like 
\emph{reading}. They can be related to concrete entities like the fonts 
used in the user interface.

\begin{figure*}[htb]
\begin{center}
  \includegraphics[width=.8\textwidth]{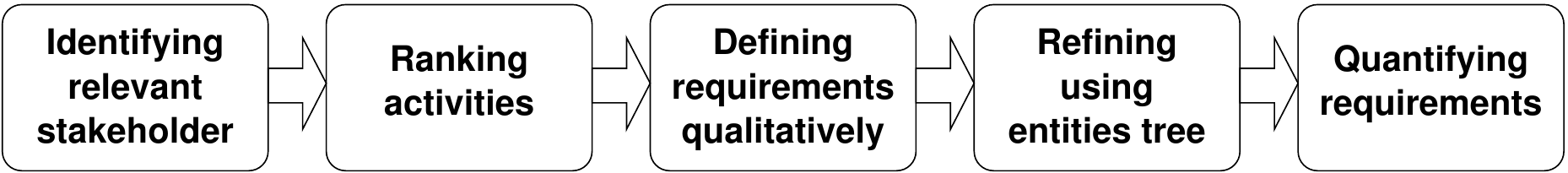}
  \caption{The process steps for quality requirements elicitation and
           refinement \label{fig:process}}
\end{center}
\end{figure*}

\section{Elicitation and Refinement}
\label{sec:elicitation}

The main approaches to elicit quality requirements are either checking
several requirements types and building
prototypes \cite{robertson99} or using positive and/or negative scenarios
(use cases and misuse cases) \cite{alexander03}. Although we believe that
both approaches are valid, important, and best used in combination, the
incorporation of the quality model described in 
Sec.~\ref{sec:quality_modelling} 
can improve the
result by defining more structure.

We use the structure induced by the quality model to elicit and refine the
quality requirements. This elicitation and refinement process consists of
5 main steps that should be supported by the established elicitation techniques 
mentioned above. An overview is shown in Fig.~\ref{fig:process}. The steps
are strongly oriented at using the two trees contained in the quality model
and aim at refining the requirements to quantitative values as far as
possible. Obviously, the approach is influenced by the availability of a
suitable quality model. Ideally, an appropriate quality model exists that
shows the needed activities and entities. However, this will often not be the
case but many activities and the upper levels of 
the entities tree can
usually be reused or found in the literature \cite{deissenboeck07}. Then the
quality model should be refined in parallel to the requirements.

\subsection{Identifying Relevant Stakeholders} The first step is, similar as
in other requirements elicitation approaches, to identify the stakeholders of
the software system. For quality requirements, this usually includes users,
developers and maintainers, operators, and user trainers. Obviously, other
stakeholders can also be relevant for the quality requirements. When the
stakeholders have been identified, the quality model can be used to derive the
activities they perform on and with the system. For example, the activities
for the maintainer include
\emph{concept location}, \emph{impact analysis},
\emph{coding}, or \emph{modification} \cite{deissenboeck07}. Especially the 
activities of the user can be further detailed by developing usage scenarios.

\subsection{Ranking Activities} In the next step, we rank the activities of 
the relevant stakeholders according to their importance. If not all needed 
activities are defined in the model, it will be extended accordingly. This 
results in a list of all activities of the relevant stakeholders. On top of 
this list are the most important activities, the least important at the 
bottom. Importance hereby means the activities that are expected to be 
performed most often and which are most elaborate. The justification can be 
given by expert opinion or experiences from similar projects. This list 
will be used in the following to focus the definition and refinement of the 
requirements.

\subsection{Defining Requirements Qualitatively} Now, we need to answer the
question how well we want the activities to be supported. The answers 
are in essence
qualitative requirements on the software system. For example, if we expect
rather complex and difficult concepts in the software because the problem
domain already contains many concepts, the activity \emph{concept location}
is desired to be \emph{simple}. These qualitative statements are needed for
all the activities. Depending on the amount of activities to be considered,
the ones at the bottom of the list might be ignored and simply judged with
\emph{don't care}.

\subsection{Refining Using the Entities Tree}
As described in Sec.~\ref{sec:quality_modelling}, the entities tree contains the
entities of the software and its environment that
are in some way relevant for the quality of the system. The entities tree
organises them in a well-defined, hierarchical manner that fits perfectly to
the task of refining the requirements elicited based on the activities. The
quality model itself is a valuable help for this. It actually captures the
influences (of properties) of entities on the activities. Hence, we only need
to follow the impacts the other way round to find the entities that have an
influence on a specific activity. If these influences are incomplete in the
current model, this step can also be used to improve it. This way, 
consistency with the later quality assurance is significantly easier to
achieve. For the definition of the refined quality requirements, the
attributes defined for the entities can be used. For example, a
detailed requirement might be that each object must have a state accessible
from outside because this has a positive effect on the \emph{test} activity.
We refine those higher-level requirements in more detail that have been
judged to be important in the last steps.

\subsection{Quantifying Requirements}
Finally, the goal is to have quantitative and hence easily 
checkable requirements.
We can quantify the requirements on the activity-level or on the entity-level.
On the activity-level, this
would be, for example, that an average \emph{modification} 
activity should take 4
person-hours to complete. Requirements on the entity-level might be
quantitatively assessable (cf.~\cite{deissenboeck07}) depending on the attribute
concerned. For example, \emph{needless code variables} can be counted and
hence an upper limit can be given.
We assume that it is theoretically possible for any requirement to define it
quantitatively. Yet, it is not always feasible in practice as either there is
no known decomposition of the requirement or it is considered too elaborate.


\section{Assurance}
\label{sec:assurance}

\begin{figure}[ht] 
\includegraphics[width=0.99\linewidth]{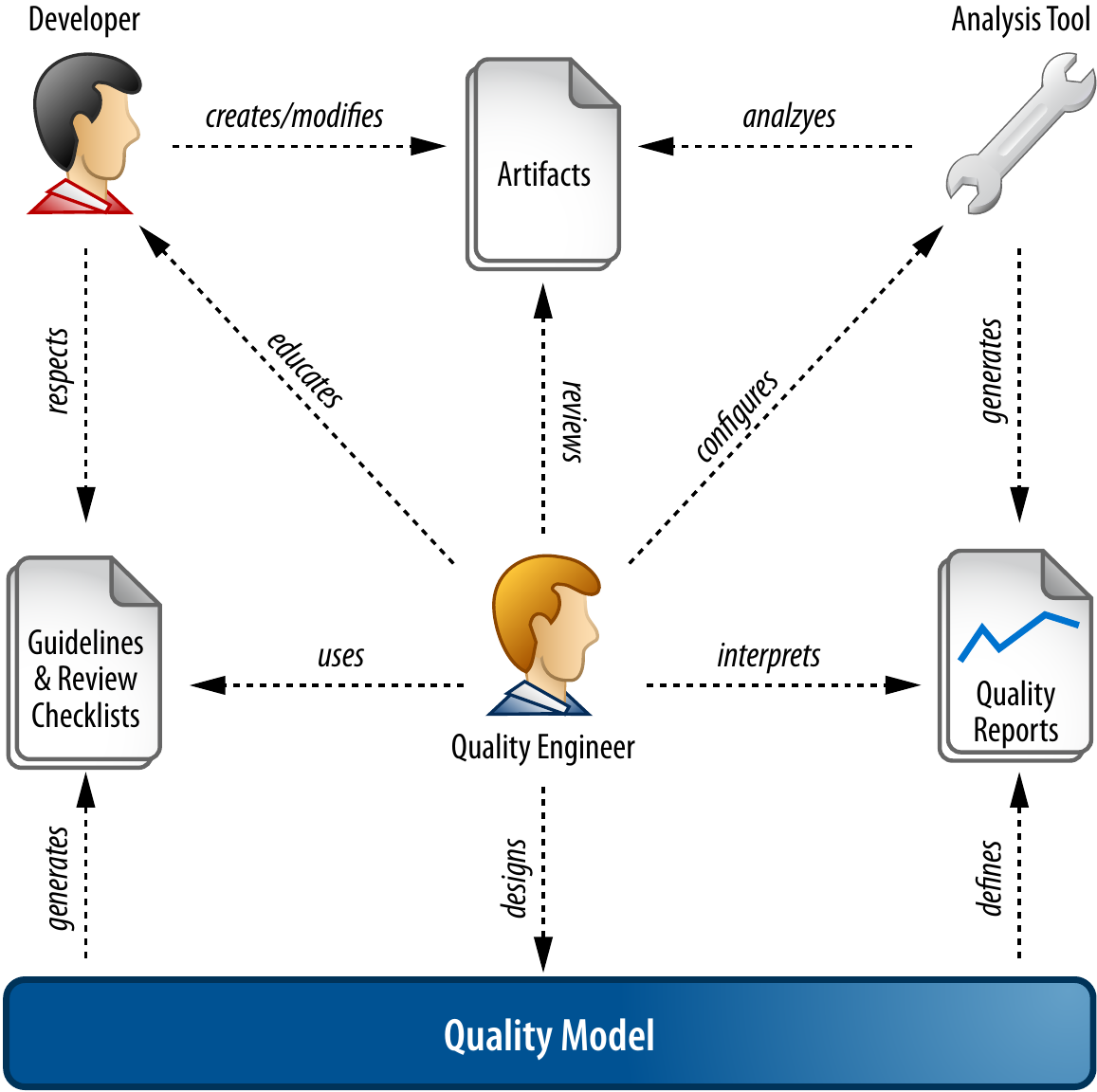} 
\caption{Model-based quality assurance \label{fig:assurance}} 
\end{figure}

The model acts as a central knowledge base for the quality-related 
relationships in the product and process. Therefore, it is also a 
well-suited basis for assuring that the defined quality requirements have 
been fulfilled. Fig.~\ref{fig:assurance} shows how the model can be used in 
several ways for constructive as well as analytical quality assurance (QA).

Constructive QA is supported by the automatic generation of \emph{Quality 
Guidelines} from the model. These guidelines define what \emph{Developers} 
should do and what they should not do in order to meet the quality 
requirements expressed by the model. For analytic QA, the \emph{Quality 
Engineer}, uses manual \emph{Reviews} as well as the \emph{Quality Reports} 
generated by \emph{Quality Analysis Tools} to evaluate if quality 
requirements are satisfied. Like the guidelines, \emph{Checklists} that 
support the manual reviews are automatically generated from the model. To 
be used efficiently, review checklists are required to be as short as 
possible~\cite{1999_brykczynskib_review_checklists}. As we annotate the 
properties in the model, whether they are automatically, semi-au\-to\-ma\-%
tic\-ally, or only manually assessable, review checklists can be limited to 
issues that require manual evaluation and thereby be kept as concise as 
possible. Checklist length can be further limited by selecting only the subset
of the quality model that is relevant for the artefact type being 
reviewed. 
Quality analysis tools like coding convention checkers or 
our assessment toolkit \textsc{ConQAT}~\cite{2005_deissenboeckf_conqat} 
report their results with respect to the entities defined in the model. 
Hence, the quality engineer can give concrete instructions to developers to 
correct quality issues. The results of quality analysis tools that require 
the execution of the system, \eg usability tests, do usually not provide 
such a direct relation to the entities. However, the model supports the 
identification of entities responsible for quality defects via the 
explicitly stored relations between entities and activities.

\section{Case Study}
\label{sec:case_study}

We show the applicability of our approach in an automotive case study.
DaimlerChrysler published a sample system specification of an
instrument cluster \cite{buhr03}. The instrument cluster is the system
behind a vehicle's dashboard controlling the rev meter, the speedometer,
indicator lights, etc. The specification is strongly focused on functional
requirements but also contains various ``business'' requirements that
consider quality aspects. The functional requirements are analysed in more
detail in \cite{kof05}. We mainly look at the software requirements but
also at how the software influences the hardware requirements.

\subsection{Identifying Relevant Stakeholders}

We can identify two stakeholders for the quality requirements stated in
\cite{buhr03}.
The relevant requirements are mainly concerned with the user of
the system, i.e., the \emph{driver}. He needs to have a good view
on all information, relevant information needs to be given directly 
and his safety
has to be ensured. To derive the corresponding activities, we can use the
quality model for usability described in \cite{winter:interactive07}. It
contains a case study about the ISO 15005 \cite{iso15005:2002} that defines
ergonomic principles for the design of transport information and control
systems (TICS). The instrument cluster is one example of such systems. 
Hence, the
identified activities can be used here. The distinction on the top level
is in \emph{driving} and \emph{TICS dialog}. The former describes the 
activity of controlling the car in order to navigate and manoeuvre it.
Examples are steering, braking, or accelerating. The latter means the
actual use of a TICS system. It is divided into: (1) \emph{view}, (2) 
\emph{perception}, (3) \emph{processing}, and
(4) \emph{input}. This level of granularity is sufficient to describe
quality related relationships \cite{winter:interactive07}.

The second important stakeholder is the manufacturer of the vehicle, the
\emph{OEM}. The concern is mainly in two directions: (1) reuse of proven 
hardware from the last series and (2) power consumption. The former is an OEM
concern because it allows decreased costs and ensures a certain level of
reliability which in turn also reduces costs by less defect fixes. The power
consumption is typically an important topic in automotive development because
of the high amount of electronic equipment that needs to be served. Hence, to
avoid a larger or a second battery -- and thereby higher costs --, the power 
consumption has to be minimised. Therefore, the relevant activities of 
the OEM are
(1) \emph{system integration} in which the software is integrated with the
hardware and (2) \emph{defect correction} which includes callbacks as well
as repairs because of warranty.

\subsection{Ranking Activities}

The above identified activities of the two relevant stakeholders need now
be ranked according to their importance for those stakeholders. The decisive
view is obviously the one from the payer, in this case the \emph{OEM}. Only
legal constraints can have a higher priority.

Although we do not know how DaimlerChrysler would prioritise these activities,
we assume that usually the safety of the driver should have the highest
priority. Hence, the \emph{driving} activity is ranked above \emph{defect
correction} and \emph{system integration}. The rationale for ranking
\emph{defect correction} higher than \emph{system integration} is that
the former is is extremely expensive, especially in case the system is in
the field already. 
This is partly backed up by the commonly known
fact that it is the more expensive to fix a defect, the later it is detected
\cite{wagner:isese06}. The complete ranking is also
shown in Tab.~\ref{tab:activities}.

\begin{figure*}[ht]
\begin{center}
  \includegraphics[width=.7\textwidth]{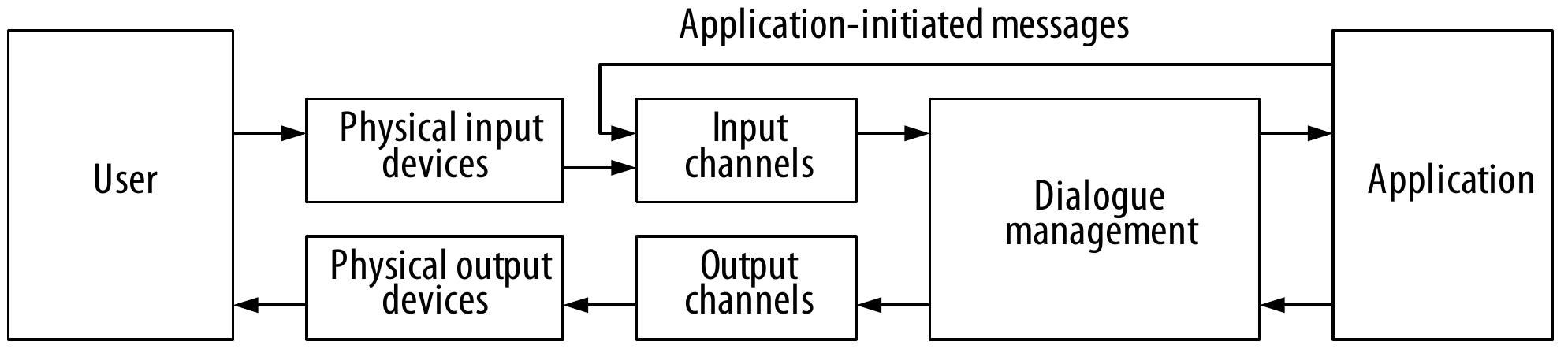}
  \caption{The abstract user interface architecture \label{fig:architecture}}
\end{center}
\end{figure*}

\subsection{Defining Requirements Qualitatively}

Having identified and prioritised the activities that are performed on
and with the system, they can be used to define the requirements qualitatively.
This way, the requirements are elicited and some requirements may not be
possible to give in finer detail. For the case study, we analyse the
``business requirements'' and their rationales (if available)
from \cite{buhr03} to derive
qualitative ratings. The ratings for the activities are summarised
in Tab.~\ref{tab:activities}.

\begin{table}[ht]
\caption{Qualitatively defined requirements for the prioritised activities
         \label{tab:activities}}
\begin{tabular}{p{2.7cm}p{5.0cm}}
\hline
\textbf{Activity} & \textbf{Rating}\\
\hline
Driving & comfortable, safe, not distracted\\
TICS Dialog & informative, attractive, correct, current, agile, dynamic,
              safe, reliable, traditional, accurate, authentic, intuitive,
              improved\\
Defect correction & minimal\\
System integration & minimal hardware requirements, using existing hardware
                     components, interoperable with different hardware\\
\hline
\end{tabular}
\end{table}
%
%
%

For all parts of the instrument cluster, it is wanted that \emph{driving}
is still comfortable. The \emph{driver} should not be distracted in the
\emph{driving} activity. Hence, it must be safe. The most information can
not surprisingly be found about the \emph{TICS dialog} itself. It is often 
stated that it should be possible to obtain information and that the dialog
should be attractive for the \emph{driver}. The information displayed needs
to be correct, current, accurate, and authentic. In general, it is also
stated that the dialog needs to be ``well-known'' what we called
``traditional'' but it must also improve over the current systems. The \emph{defect correction}
should be minimal with a high robustness and life-span. Finally, \emph{system
integration} should have minimal hardware requirements and use existing
hardware components. It should also be able to use different hardware,
especially in the case of different radio vendors.

\subsection{Refining Using Entities Tree}

We can again use the quality model from \cite{winter:interactive07} for
most of the entities tree. It provides a decomposition of the \emph{vehicle}
into the \emph{driver} and \emph{TICS}. The \emph{TICS} is further divided
into \emph{hardware} and \emph{software}. The software is decomposed
based on an abstract
architecture of user interfaces from \cite{winter:interactive07}
as depicted in Fig.~\ref{fig:architecture}. The hardware is divided into
operating devices, indicators/display, and the actual TICS unit.

The quality model gives us also the connection from activities to those
entities. It shows which entities need to be considered w.r.t.~the
activities of the stakeholders that are important for our instrument cluster.
We cannot describe this completely for reasons of brevity but give some
examples for refinements using the entities tree.

For the \emph{driving} activity, we have a documented influence from the
the hardware, for example. More specifically, the appropriateness
of the position of the  
display has an influence on \emph{driving}. In our more formal
notation that is:

\begin{center}
\impact{$\mathsf{Display.Position}$}{$\mathsf{Appropriateness}$}
{$\mathsf{Driving}$}{+}
\end{center}

Hence, in order to reach the qualitative goals for the \emph{driving}
activity, we need to ensure that the display position is appropriate.

A second example starts from the \emph{processing} activity. It is
influenced by the unambiguousness of the representation of the output
data:

\begin{center}
\impact{$\mathsf{Output Data.Representation}$}{$\mathsf{Unambigousness}$}
{$\mathsf{TICS dialog.Processing}$}{+}
\end{center}

Therefore, we have a requirement on the representation of the output
data that it must be unambiguous, i.e., the driver understands the
priority of the information.

Finally, \emph{perception} is an activity in the \emph{TICS dialog} that
is influenced by the adaptability of the output data representation:

\begin{center}
\impact{$\mathsf{Output Data.Representation}$}{$\mathsf{Adaptability}$}
{$\mathsf{TICS dialog.Perception}$}{+}
\end{center}

The representation should be adapted to different driving situations
so that the time for \emph{perception} is minimised. Such a requirement
is currently missing in the specification \cite{buhr03}.

\subsection{Quantifying Requirements}

For the quantification of the requirements, the quality model can only
help if there are metrics defined for measuring the facts. Then an
appropriate value can be defined for that metric. Otherwise, the model
must be extended here with a metric, if possible. We can again not
describe all necessary quantifications of the instrument cluster
specification but provide some examples.

The above identified requirement about the appropriateness of the display
position can be given a quantification. The specification \cite{buhr03}
actually demands that ``The display tolerance [\ldots] amounts to
$\pm 1.5$ degrees.'' Furthermore, it is stated that ``The angle of deflection
of the pointer of the rev meter display amounts to 162 degrees.''

The example of the unambiguous representation of the output data cannot
be described with some kind of numerical value. However, the specification
\cite{buhr03} demands that the engine control light must not be placed
in the digital display with lots of other information ``because an own
place in the instrument cluster increases its importance''.

\subsection{Discussion}

For reasons of brevity, we are not able to describe the whole quality
requirements elicitation and refinement for the instrument cluster.
However, the examples show that our approach is applicable to
such an automotive system. We observed that we have a clear guidance
in eliciting and refining the requirements along the quality model.
Starting from the stakeholders, their activities down to the influencing
system entities is a straight-forward thinking process. Moreover, we
found that several of the informations needed during the application
of our approach was already contained in the specification but not consistently
for all its parts. Finally, we also noted that we were able to identify
several requirements that were not considered in the specification that
can have an influence on the relevant activities.

\section{Conclusions}
\label{sec:conclusions}

Although it has been acknowledged that quality requirements are difficult to
handle and that they often have been neglected, a well-founded and agreed 
structuring of those requirements has not been established. In some way, most
classifications are related to the ISO 9126 standard
\cite{2003_iso_standard_9126_1} that, however, does not consider the various
activities of the stakeholders. Our unique, two-dimensional quality model
\cite{deissenboeck07} resolves this problem by explicitly modelling the
influences of entities on activities. Using the
activities of the stakeholders and following these impacts in the opposite
direction, we can employ a structured and well-founded process to elicit and
refine the quality requirements. 

Moreover, there is a direct connection and traceability to the 
quality assurance
techniques used in the project. Those techniques are responsible for assuring
that the quality requirements have been fulfilled. The quality model serves
as a basis of the quality assurance and thereby allows to relate the results
to the requirements.

The applicability of the approach was shown in a case study based on a published
instrument cluster specification of DaimlerChrysler. We were able to show
that the approach allows a structured elicitation and refinement of quality
requirements using the activities of the stakeholders and their relationships
with entities in the system that are documented in a quality model. We found
information that could be fitted into our approach but also could
identify omissions in the specification.

We plan to develop the approach in more detail and to validate it in more
case studies. Moreover, the quality model itself is continuously extended and
improved. This in turn also amends the quality requirements approach.

\section{Acknowledgements}
We are grateful to K.~Buhr, N.~Heumesser, F.~Houdek, H.~Omasreiter, 
F.~Rothermel,
R.~Tavakoli, and T.~Zink for the specification of the instrument cluster
and D. Mendez for useful comments.
\balance

\end{document}